\documentclass[aps,twocolumn,prd,superscriptaddress,preprintnumbers,showpacs]{revtex4-1}
\usepackage{graphicx}
\usepackage{bm}
\usepackage{amsmath}
\usepackage{amssymb}
\usepackage{amsthm}

\newcommand{\be}{\begin{equation}}
\newcommand{\ee}{\end{equation}}
\newcommand{\bea}{\begin{eqnarray}}
\newcommand{\eea}{\end{eqnarray}}
\newcommand{\bdm}{\begin{displaymath}}
\newcommand{\edm}{\end{displaymath}}
\newcommand{\beas}{\begin{eqnarray*}}
\newcommand{\eeas}{\end{eqnarray*}}

\begin{document}
\title{Consistency tests of the stability of fundamental couplings and unification scenarios}

\author{M. C. Ferreira}
\email[]{up080302013@alunos.fc.up.pt}
\affiliation{Centro de Astrof\'{\i}sica, Universidade do Porto, Rua das Estrelas, 4150-762 Porto, Portugal}
\affiliation{Faculdade de Ci\^encias, Universidade do Porto, Rua do Campo Alegre, 4150-007 Porto, Portugal}
\author{O. Frigola}
\affiliation{Institut S'Agulla, Carretera de Malgrat 13, 17300 Blanes, Spain}
\author{C. J. A. P. Martins}
\email[]{Carlos.Martins@astro.up.pt}
\affiliation{Centro de Astrof\'{\i}sica, Universidade do Porto, Rua das Estrelas, 4150-762 Porto, Portugal}
\author{A. M. R. V. L. Monteiro}
\email[]{mmonteiro@fc.up.pt}
\affiliation{Centro de Astrof\'{\i}sica, Universidade do Porto, Rua das Estrelas, 4150-762 Porto, Portugal}
\affiliation{Faculdade de Ci\^encias, Universidade do Porto, Rua do Campo Alegre, 4150-007 Porto, Portugal}
\author{J. Sol\`a}
\affiliation{Institut Manuel Blancafort, Avinguda 11 de Setembre 29, 08530 La Garriga, Spain}

\date{2 April 2014}

\begin{abstract}
We test the consistency of several independent astrophysical measurements of fundamental dimensionless constants. In particular, we compare direct measurements of the fine-structure constant $\alpha$ and the proton-to-electron mass ratio $\mu=m_p/m_e$ (mostly in the optical/ultraviolet) with combined measurements of $\alpha$, $\mu$ and the proton gyromagnetic ratio $g_p$ (mostly in the radio band). We point out some apparent inconsistencies, which suggest that hidden systematics may be affecting some of the measurements. These findings demonstrate the importance of future more precise measurements with ALMA, ESPRESSO and ELT-HIRES. We also highlight some of the implications of the currently available measurements for fundamental physics, specifically for unification scenarios.
\end{abstract}

\maketitle

%%%%%%%%%%%%%%%%%%%%%%%%%%%%%%%%%%%%%%%%%%%%%%%%%%%%%%%%%%%%%%%%%%%%%%%%%%%%%%
\section{\label{intro}Introduction} 

The stability of nature's fundamental couplings is among the most profound open issues in astrophysics and fundamental physics, and has been identified by the European Space Agency (ESA) and the European Southern Observatory (ESO) as one of the key drivers for the next generation of ground and space-based facilities. While historically we have assumed that these are spacetime-invariant, this is only a simplifying assumption, whose only possible justification is an appeal to Occam's razor.

Although we have no 'theory of couplings', that describes their role in physical theories (or even which of them are really fundamental), at a phenomenological level it is well known that fundamental couplings \textit{run} with energy, and in many extensions of the standard model they will also \textit{roll} in time and \textit{ramble} in space (ie, they will depend on the local environment). In particular, this will be the case in theories with additional spacetime dimensions, such as string theory. A detection of varying fundamental couplings will be revolutionary: it will automatically prove that the Einstein Equivalence Principle is violated and that there is a fifth force of nature. Reviews of the subject can be found in \cite{RSoc,GBerro,Uzan}.

One must also realize that even improved null results are important. Naively, the natural scale for cosmological evolution of one of these couplings (if one assumes that it is driven by a scalar field) would the Hubble time. We would therefore expect a drift rate of the order of $10^{-10}$ yr${}^{-1}$. However, current local bounds coming from atomic clock comparison experiments \cite{Rosenband}, are already about 6 orders of magnitude stronger, and rule out many otherwise viable dynamical dark energy models. A recent combined analysis of all currently available atomic clock measurements \cite{Clocks} led to the following indirect 95\% confidence intervals for present-day ($z=0$) drift rates
\begin{equation}
\frac{\dot\mu}{\mu}=(6.8\pm57.6)\times10^{-17}\,{\rm yr}{}^{-1}\,
\end{equation}
\begin{equation}
\frac{\dot g_p}{g_p}=(-7.2\pm8.9)\times10^{-17}\,{\rm yr}{}^{-1}\,,
\end{equation}
to be compared to the direct experimental result of \cite{Rosenband} for the fine-structure constant (also at the 95\% confidence level)
\begin{equation}
\frac{\dot \alpha}{\alpha}=(-1.7\pm4.9)\times10^{-17}\,{\rm yr}{}^{-1}\,.
\end{equation}

Astrophysical measurements have led to claims for \cite{Webb,Reinhold,Dipole} and against \cite{Chand,King,Thompson} variations of the fine-structure constant $\alpha=e^2/\hbar c$ and the proton-to-electron mass ratio $\mu=m_p/m_e$ at redshifts $z\sim1-3$. An ongoing Large Program at European Southern Observatory's Very Large Telescope (henceforth referred to as the LP) is expected to clarify matters \cite{LP1,LP2}, but a resolution may have to wait for a forthcoming generation of high-resolution ultra-stable spectrographs such as ESPRESSO \cite{Respresso} and ELT-HIRES \cite{Rhires1,Rhires2} which include improving these measurements among their key science goals (and in some cases this actually drives the instrument's design). Answering this question is also essential in order to shed light on the enigma of dark energy \cite{Parkinson,NunLid,Doran,Reconst}.

Any Grand-Unified model predicts a specific relation between the variation of $\alpha$ and those of $\mu$ and other couplings, and therefore simultaneous measurements of both provide key consistency tests, with direct implications for the phenomenology of these models. The basic formalism for these tests was developed in \cite{Coc,Luo} (simpler, and somewhat more model-dependent studies were also done in \cite{DSW,Berengut}). This formalism has subsequently been applied to astrophysical observations of solar-type stars \cite{Vieira} and neutron stars \cite{Angeles}, as well as to local experiments with atomic clocks \cite{Clocks}. More recently we have also applied this formalism to observations of the radio source PKS1413$+$135 \cite{PKS}.

Here, we extend these previous analyses to a much broader set of astrophysical measurements of fundamental couplings. These measurements span a wide range of redshifts and were obtained with an equally diverse range of observational techniques, from the optical/UV to the radio band. We will simply take the various (independent) available measurements at face value and check for their internal consistency. Unless otherwise is stated, the results of these spectroscopic measurements will be presented in units of parts per million (ppm), ie. $10^{-6}$.

We note that in the present analysis we will be focusing on time variations, and for that reason we do not explicitly include the results of Webb \textit{et al.} \cite{Dipole}, which provide some evidence for spatial variations, in our analysis. Extending our formalism to allow for possible spatial variations is possible in principle but it requires additional statistical analysis tools; for this reason, it is left for subsequent work.

%%%%%%%%%%%%%%%%%%%%%%%%%%%%%%%%%%%%%%%%%%%%%%%%%%%%%%%%%%%%%%%%%%%%%%%%%%%%%%
\section{\label{unify}Phenomenological models}

We shall work on the assumption that varying fundamental couplings are due to a dynamical, dilaton-type scalar field. In this section we describe the specific class of unification scenarios we will be considering, as well as the possible effects of couplings between this degree of freedom and photons (which can have relevant observational consequences). The required theoretical formalism has been described in our previous papers \cite{Clocks,PKS,Tofz1,Tofz2} and in other references therein; here we will simply provide a brief summary.

\subsection{Unification scenarios}

We consider a class of grand unification models (in which unification happens at an unspecified high-energy scale) where the weak scale is determined by dimensional transmutation and the relative variation of all the Yukawa couplings is the same. We also assume that the variation of the couplings is driven by a dilaton-type scalar field (as in \cite{Campbell}).

With these assumptions one finds that the variations of $\mu$ and $\alpha$ are related via
\begin{equation}
\frac{\Delta\mu}{\mu}=[0.8R-0.3(1+S)]\frac{\Delta\alpha}{\alpha}\,,
\end{equation}
where $R$ and $S$ are phenomenological (model-dependent) parameters, whose absolute values can be anything from order unity to several hundreds. Although physically one may expect them to be positive, for our purposes they can be taken as free parameters to be constrained by data. At a phenomenological level, the choice $S=-1$, $R=0$ can also describe the limiting case where $\alpha$ varies but the masses don't.

Further useful relations can be obtained \cite{flambaum1,flambaum2} for the gyromagnetic ratios of the proton and neutron
\begin{equation}
\frac{\Delta g_p}{g_p}=[0.10R-0.04(1+S)]\frac{\Delta\alpha}{\alpha}\,
\end{equation}
\begin{equation}
\frac{\Delta g_n}{g_n}=[0.12R-0.05(1+S)]\frac{\Delta\alpha}{\alpha}\,.
\end{equation}

These allow us to transform any measurement of a combination of $\alpha$, $\mu$ and $g_p$ into a constraint on the $(R,S,\alpha)$ parameter space. For atomic clocks, the relevant g-factors are those for Rubidium and Caesium, which can be related to those of the nucleons \cite{Luo,Clocks}. 

\subsection{CMB temperature evolution}

Overviews of the possible effects of scalar fields on the redshift evolution of the CMB temperature can be found in \cite{Tofz1,Tofz2}. If there is a coupling between the scalar field and the radiation fluid, the photon temperature-redshift relation will be distorted away from its standard evolution. At a phenomenological level, this can be described with the additional parameter
\begin{equation}
T(z)=T_0(1+z)^{1-\beta}\,,
\end{equation}
with the available measurements of $T(z)$ providing percent-level constraints $\beta$ \cite{Luzzi,Noterdaeme,Tofz1}. The corresponding evolution of the radiation density is
\begin{equation}
\rho_\gamma\propto T^4\propto (1+z)^{4(1-\beta)}\propto a^{-4(1-\beta)}\,,
\end{equation}
with $a$ being the cosmological scale factor.

One specific example of such a class of phenomenological models is the Bekenstein-Sandvik-Barrow-Magueijo (BSBM) \cite{Sandvik}, for which it has been shown \cite{Tofz2} that
\begin{equation}
\frac{T(z)}{T_0}=(1+z)\left(\frac{\alpha(z)}{\alpha_0}\right)^{1/4}\sim(1+z)\left(1+\frac{1}{4}\frac{\Delta\alpha}{\alpha}\right)\,.
\end{equation}
In this specific case we have $\alpha$ variations without corresponding variations of $\mu$ (as previously pointed out, in our formalism this can be described by the parameter choice $R=0$, $S=-1$). But regardless of the BSBM specific case, we can similarly take this as a phenomenological relation that can be tested observationally. There is currently no system for which both $T(z)$ and $\alpha$ have been measured, but such systems do exist for $\mu$. In particular, \cite{Salumbides} points out that the $A^1\Pi-X^1\Sigma^+$ band system of CO, which has been detected in six galaxies in the redshift range $1.6$--$2.7$, is a probe method for $\mu$, while as already demonstrated in \cite{Noterdaeme} CO is also ideal to measure $T(z)$.

It is convenient to define the relative temperature variation
\begin{equation}\label{deltaT}
\frac{\Delta T(z)}{T}=\frac{T_{\rm obs}(z)-T_{\rm std}}{T_{\rm std}}=\frac{T_{\rm obs}(z)}{T_0(1+z)}-1\,
\end{equation}
with the local ($z=0$) measurement being \cite{Mather}
\begin{equation}
T_0=(2.725\pm0.002)\,K\,.
\end{equation}
With these definitions and the above results we have
\begin{equation}
\frac{\Delta T(z)}{T}=\frac{1}{4}\frac{\Delta\alpha}{\alpha}\,
\end{equation}
and therefore
\begin{equation}
\frac{\Delta\mu}{\mu}=4[0.8R-0.3(1+S)]\frac{\Delta T(z)}{T}\,
\end{equation}
which provides a further consistency test.

%%%%%%%%%%%%%%%%%%%%%%%%%%%%%%%%%%%%%%%%%%%%%%%%%%%%%%%%%%%%%%%%%%%%%%%%%%%%%%
\section{\label{qsodata}Current spectroscopic measurements}

In this section we list the current astrophysical measurements that we will be using in our analysis. (As previously mentioned, we'll usually list them in units of parts per million.) This is not meant to be a 'historical' review listing all available measurements. In most cases we use only the tightest available measurement for each astrophysical source. A few older measurements along other lines of sight have not been used, on the grounds that they would have no statistical weight in our analysis. Nevertheless, we will include some low-sensitivity but high-redshift measurements, as these are illustrative of the redshift range that may be probed by future facilities. Our two exceptions regarding measurements of the same source concern
\begin{itemize}
\item Measurements using different (that is, independent) techniques---typically, measurements of $\mu$ or combined measurements using different molecules, and
\item Measurements obtained with different spectrographs\,.
\end{itemize}

\begin{table}
\begin{tabular}{|c|c|c|c|c|}
\hline
Object & z & $Q_{AB}$  & ${ \Delta Q_{AB}}/{Q_{AB}}$ & Ref. \\ 
\hline\hline
PKS1413$+$135 & 0.247 & ${\alpha^{2\times1.85}g_{p}\mu^{1.85}}$  & $-11.8\pm4.6$ & \protect\cite{Kanekar2} \\
\hline
PKS1413$+$135 & 0.247 & ${\alpha^{2\times1.57}g_{p}\mu^{1.57}}$  & $5.1\pm12.6$ & \protect\cite{Darling} \\
\hline
PKS1413$+$135 & 0.247 & ${\alpha^{2}g_{p}}$  & $-2.0\pm4.4$ & \protect\cite{Murphy} \\
\hline\hline
B0218$+$357 & 0.685 & ${\alpha^{2}g_{p}}$ & $-1.6\pm5.4$ & \protect\cite{Murphy} \\
\hline\hline
J0134$-$0931 & 0.765 & ${\alpha^{2\times1.57}g_{p}\mu^{1.57}}$  &  $-5.2\pm4.3$ & \protect\cite{Kanekar} \\
\hline\hline
J2358$-$1020 & 1.173 & ${\alpha^{2}g_{p}/\mu}$ & $1.8\pm2.7$ & \protect\cite{Rahmani} \\
\hline\hline
J1623$+$0718 & 1.336 & ${\alpha^{2}g_{p}/\mu}$ & $-3.7\pm3.4$ & \protect\cite{Rahmani} \\
\hline\hline
J2340$-$0053 & 1.361 & ${\alpha^{2}g_{p}/\mu}$ & $-1.3\pm2.0$ & \protect\cite{Rahmani} \\
\hline\hline
J0501$-$0159 & 1.561 & ${\alpha^{2}g_{p}/\mu}$ & $3.0\pm3.1$ & \protect\cite{Rahmani} \\
\hline\hline
J0911$+$0551 & 2.796 & ${\alpha^{2}\mu}$ & $ -6.9\pm3.7$ & \protect\cite{Weiss} \\
\hline\hline
J1337$+$3152  & 3.174 & ${\alpha^{2}g_{p}/\mu}$ & $-1.7\pm1.7$ & \protect\cite{Petitjean1} \\
\hline\hline
BR1202$-$0725 & 4.695 & ${\alpha^{2}\mu}$ & $50\pm150$ & \protect\cite{Lentati} \\
\hline\hline
J0918$+$5142 & 5.245 & ${\alpha^{2}\mu}$ & $-1.7\pm8.5$ & \protect\cite{Levshakov} \\
\hline\hline
J1148$+$5251 & 6.420 & ${\alpha^{2}\mu}$ & $330\pm250$ & \protect\cite{Lentati} \\
\hline\hline\hline
\end{tabular}
\caption{\label{table1}Available measurements of several combinations of the dimensionless couplings $\alpha$, $\mu$ and $g_p$. Listed are, respectively, the object along each line of sight, the redshift of the measurement, the dimensionless parameter being constrained, the measurement itself (in parts per million), and its original reference.}
\end{table}

Table \ref{table1} contains current joint measurements of several couplings. Note that for the radio source PKS1413$+$135 the three available measurements are sufficient to yield individual constraints on the variations of the three quantities at redshift $z=0.247$. This was done in \cite{PKS}, which at the one-sigma (68.3$\%$) confidence level obtained
\begin{equation}\label{pkslike1}
\frac{\Delta \alpha}{\alpha}=-51\pm43\, {\rm ppm}\,
\end{equation}
\begin{equation}\label{pkslike2}
\frac{\Delta\mu}{\mu}=41\pm39\, {\rm ppm}\,
\end{equation}
\begin{equation}\label{pkslike3}
\frac{\Delta g_p}{g_p}=99\pm86 \, {\rm ppm}\,,
\end{equation}

\begin{table}
\begin{tabular}{|c|c|c|c|c|}
\hline
 Object & z & ${ \Delta\alpha}/{\alpha}$ & Spectrograph & Ref. \\ 
\hline\hline
HE0515$-$4414 & 1.15 & $-0.1\pm1.8$ & UVES & \protect\cite{alphaMolaro} \\
\hline
HE0515$-$4414 & 1.15 & $0.5\pm2.4$ & HARPS/UVES & \protect\cite{alphaChand} \\
\hline
HE0001$-$2340 & 1.58 & $-1.5\pm2.6$ &  UVES & \protect\cite{alphaAgafonova}\\
\hline
{\bf HE2217$-$2818} & {\bf 1.69} & \bf {$1.3\pm2.6$} &  {\bf UVES-LP} & {\bf \protect\cite{LP1}}\\
\hline
Q1101$-$264 & 1.84 & $5.7\pm2.7$ &  UVES & \protect\cite{alphaMolaro}\\
\hline\hline\hline
\end{tabular}
\caption{\label{table2}Available specific measurements of $\alpha$. Listed are, respectively, the object along each line of sight, the redshift of the measurement, the measurement itself (in parts per million), the spectrograph, and the original reference. The entry in bold corresponds to the recent LP measurement.}
\end{table}

Table \ref{table2} contains individual $\alpha$ measurements. Conservatively we only list measurements where data was acquired specifically for this purpose---but these are, in many cases, the ones with the smallest uncertainties. In particular, the measurement listed in bold, towards HE2217$-$2818, comes from the ongoing LP and is, arguably, the one with the most robust control of systematics (the quoted error bar includes both statistical and systematic uncertainties, added in quadrature).

\begin{table}
\begin{tabular}{|c|c|c|c|c|}
\hline
 Object & z & ${\Delta\mu}/{\mu}$ & Method & Ref. \\ 
\hline\hline
B0218$+$357 & 0.685 & $0.74\pm0.89$ & $NH_3$/$HCO^+$/$HCN$ & \protect\cite{Murphy2} \\
\hline
B0218$+$357 & 0.685 & $-0.35\pm0.12$ & $NH_3$/$CS$/$H_2CO$ & \protect\cite{Kanekar3} \\
\hline\hline
PKS1830$-$211 & 0.886 & $0.08\pm0.47$ &  $NH_3$/$HC_3N$ & \protect\cite{Henkel}\\
\hline
PKS1830$-$211 & 0.886 & $-1.2\pm4.5$ &  $CH_3NH_2$ & \protect\cite{Ilyushin}\\
\hline
PKS1830$-$211 & 0.886 & $-2.04\pm0.74$ & $NH_3$ & \protect\cite{Muller}\\
\hline
PKS1830$-$211 & 0.886 & $-0.001\pm0.103$ &  $CH_3OH$ & \protect\cite{Bagdonaite2}\\
\hline\hline
J2123$-$005 & 2.059 & $8.5\pm4.2$ & $H_2$/$HD$ (VLT)& \protect\cite{vanWeerd} \\
\hline
J2123$-$005 & 2.059 & $5.6\pm6.2$ & $H_2$/$HD$ (Keck)& \protect\cite{Malec} \\
\hline\hline
{\bf HE0027$-$1836} & {\bf 2.402} & {\bf $-7.6\pm10.2$} & {\bf $H_2$} & {\bf \protect\cite{LP2}} \\
\hline\hline
Q2348$-$011 & 2.426 & $-6.8\pm27.8$ & $H_2$ & \protect\cite{Bagdonaite} \\
\hline\hline
Q0405$-$443 & 2.597 & $10.1\pm6.2$ & $H_2$ & \protect\cite{King} \\
\hline\hline
J0643$-$504 & 2.659 & $7.4\pm6.7$ & $H_2$ & \protect\cite{Albornoz} \\
\hline\hline
Q0528$-$250 & 2.811 & $0.3\pm3.7$ & $H_2$/$HD$ & \protect\cite{King2} \\
\hline\hline
Q0347$-$383 & 3.025 & $2.1\pm6.0$ & $H_2$ & \protect\cite{Wendt} \\
\hline\hline\hline
\end{tabular}
\caption{\label{table3}Available measurements of $\mu$. Listed are, respectively, the object along each line of sight, the redshift of the measurement, the measurement itself, the molecule(s) used, and the original reference. The entry in bold corresponds to the recent LP measurement.}
\end{table}

Table \ref{table3} contains individual $\mu$ measurements. Note that several different molecules can be used (with ammonia being the most common at low redshift and molecular hydrogen at high redshift), and in the case of the gravitational lens PKS1830$-$211 several independent measurements exist. The tightest available constraint was obtained in this system, from observations of methanol transitions \cite{Bagdonaite2}. On the other hand, two very recent publications containing measurements along the line of sight towards J0643$-$504 both suggest variations, though at different levels of significance. We have listed the more conservative one, from \cite{Albornoz} on the table; for comparison \cite{BagdonaiteNew} finds $\Delta\mu/\mu=17.1\pm5.0$ ppm. The measurement in bold, towards HE0027$-$1836, again comes from the ongoing LP.

\begin{table*}
\begin{tabular}{|c|c|c|c|c|c|c|}
\hline
 Object & z & ${\Delta\mu/\mu}$ (ppm) & Ref. & $T_{\mathrm{CMB}}$ (K)  & Ref. &  ${\Delta T/T}$\\ 
\hline\hline
PKS1830-211 & 0.886 & $-0.001\pm0.103$ & \protect\cite{Bagdonaite2} & ${5.08\pm0.10}$ & \protect\cite{MullerT} & $-0.012\pm0.019$\\
\hline
Q0347-383 & 3.025 & $2.1\pm6.0$ & \protect\cite{Wendt} & ${12.1^{+1.7}_{-3.2}}$ & \protect\cite{MolaroT} & $0.10^{+0.15}_{-0.29}$\\
\hline\hline
\end{tabular}
\caption{\label{table4}System with both T(z) and $\mu$ measurements. Listed are, respectively, the object along each line of sight, the redshift of the measurement, the $\mu$ and T(z) measurements with the corresponding references, and the derived value of the relative temperature variation.}
\end{table*}

Finally, Table \ref{table4} contains the systems with both $\mu$ and $T(z)$ measurements, as well as the corresponding relative temperature variation defined in Eq. (\ref{deltaT}). Clearly, these measurements have relatively large uncertainties and therefore currently have no significant additional constraining power. However, we include them here as a proof of concept, since the significantly better sensitivity of ALMA and ELT-HIRES is expected to improve both the quantity and the accuracy of these measurements.

%%%%%%%%%%%%%%%%%%%%%%%%%%%%%%%%%%%%%%%%%%%%%%%%%%%%%%%%%%%%%%%%%%%%%%%%%%%%%%
\section{\label{res}Analysis}

We now proceed to describe several different consistency studies. In each case we will mainly focus on the consistency of each set of measurements, but we will also provide a brief discussion of what these correspond to in terms of constraints in the phenomenological $R-S$ unification plane.

\subsection{Self-consistency}

%%%%%%%%%%%%%%%%%%%%%%%%%%%%%%%%%%%%%%%%%%%%%%%%%%%%%%%%%
\begin{figure}
\begin{center}
\includegraphics[width=9cm]{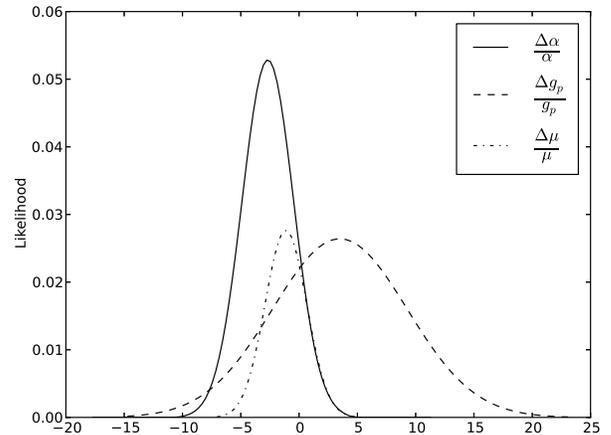}
\end{center}
\caption{One-dimensional relative likelihoods for the relative variations of $\alpha$, $\mu$ and $g_p$, from the data in Table \protect\ref{table1}.}
\label{fulltable}
\end{figure}
%%%%%%%%%%%%%%%%%%%%%%%%%%%%%%%%%%%%%%%%%%%%%%%%%%%%%%%%%

We will start by obtaining the individual values for $\alpha$, $\mu$ and $g_p$ derived from the measurements in Table \ref{table1}. In doing this we will be neglecting a possible redshift-dependence of the variations; this is of course an approximation, but given the small number of available measurements (and the large uncertainties in some of them) it is a legitimate exercise. With better data the exercise could be separately done for different redshift bins. Indeed, the present analysis is an extension of a previous analysis \cite{PKS} where we only looked at the three measurements at $z=0.247$. Figure \ref{fulltable} shows the 1D likelihood contours for each of these parameters. At the one-sigma ($68.3\%$) confidence level we find
\begin{equation}\label{fullt1}
\frac{\Delta \alpha}{\alpha}=-2.7\pm2.2\, {\rm ppm}
\end{equation}
\begin{equation}\label{fullt2}
\frac{\Delta \mu}{\mu}=-1.1\pm1.8\, {\rm ppm}
\end{equation}
\begin{equation}\label{fullt3}
\frac{\Delta g_p}{g_p}=3.5\pm5.8\, {\rm ppm}
\end{equation}
and at the two-sigma level all are consistent with no variations. Notice the significantly smaller uncertainties as compared to the results of Eq. (\ref{pkslike1}--\ref{pkslike3}) for the source at $z=0.247$ alone: the one-sigma uncertainties on $\alpha$, $\mu$ and $g_p$ are respectively reduced by factors of about 20, 22 and 15. Cpmparing these results with those listed in Tables \ref{table2} and \ref{table3} we also note that the values of $\alpha$ and $\mu$ are consistent with those found by the LP, but not with some of the other measurements in Tables \ref{table2}-\ref{table3} (or with \cite{BagdonaiteNew}).

%%%%%%%%%%%%%%%%%%%%%%%%%%%%%%%%%%%%%%%%%%%%%%%%%%%%%%%%%
\begin{figure}
\begin{center}
\includegraphics[width=9cm]{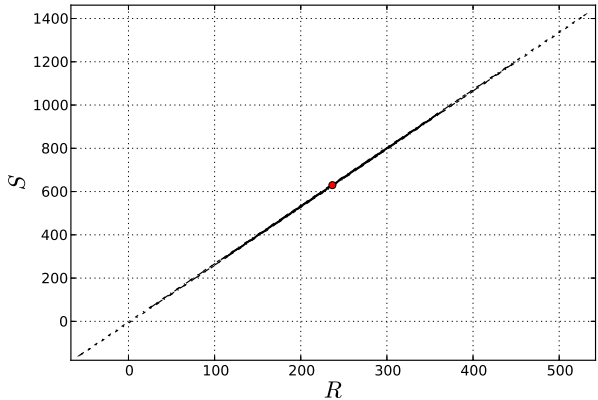}
\includegraphics[width=9cm]{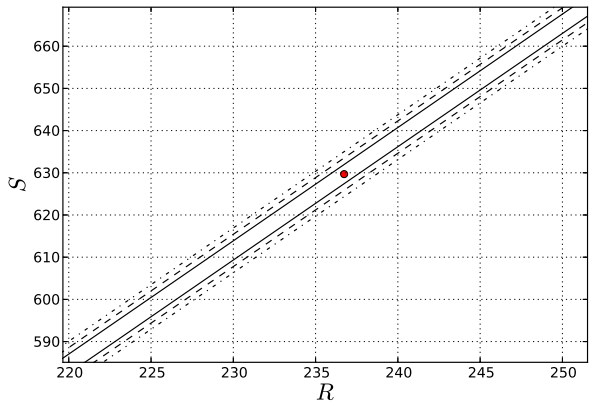}
\end{center}
\caption{Constraints on the $R$-$S$ plane from the data in Table \protect\ref{table1}. Solid, dashed and dotted lines correspond to one, two and three sigma contours. The bottom panel is a zoomed version of the top one, around the best-fit value.}
\label{fulltableRS}
\end{figure}
%%%%%%%%%%%%%%%%%%%%%%%%%%%%%%%%%%%%%%%%%%%%%%%%%%%%%%%%%

By using the relations discussed in Sect. \ref{unify} we can also translate the measurements of Table \ref{table1} into constraints on the $R-S$ plane. These are shown in Fig. \ref{fulltableRS}. As in previous analyses \cite{Clocks,PKS} there is a clear degeneracy direction: in other words, these measurements constrain a particular combination of the phenomenological parameters $R$ and $S$. It's also straightforward to determine the one-dimensional confidence intervals for $R$ and $S$; at the one-sigma confidence level we find
\begin{equation}
R=237\pm86
\end{equation}
\begin{equation}
S=630\pm230\,;
\end{equation}
these are fairly similar to the ones we found in \cite{PKS} for the three measurements towards PKS1413$+$135, showing that these carry a significant weight in the overall analysis.

\subsection{Additional $\alpha$ measurements}

The data in table \ref{table1} can also be analyzed assuming one of the direct $\alpha$ measurements in Table \ref{table2}. This will provide constraints on the $\mu-g_p$ plane as well as on the $R-S$ plane.

%%%%%%%%%%%%%%%%%%%%%%%%%%%%%%%%%%%%%%%%%%%%%%%%%%%%%%%%%
\begin{figure}
\begin{center}
\includegraphics[width=9cm]{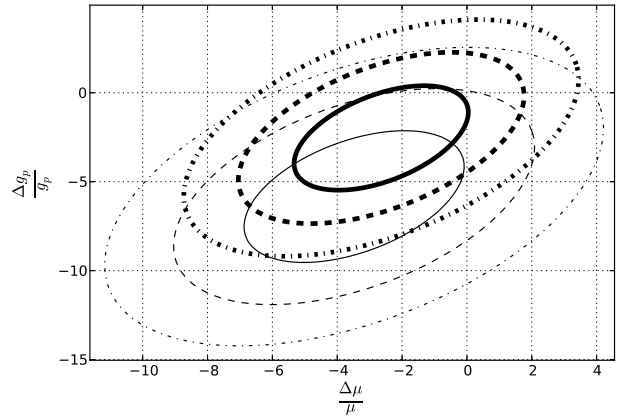}
\end{center}
\caption{One, two and three sigma constraints on the $\mu-g_p$ plane, assuming a given $\alpha$ measurement in Table \protect\ref{table2}. The thick (top) set of contours uses the $\alpha$ measurement of \protect\cite{alphaMolaro} (nominally the one with the smallest uncertainty), while the thin (bottom) set uses the LP one \protect\cite{LP1}; both axis are in ppm units.}
\label{combalpha1}
\end{figure}
%%%%%%%%%%%%%%%%%%%%%%%%%%%%%%%%%%%%%%%%%%%%%%%%%%%%%%%%%
\begin{figure}
\begin{center}
\includegraphics[width=9cm]{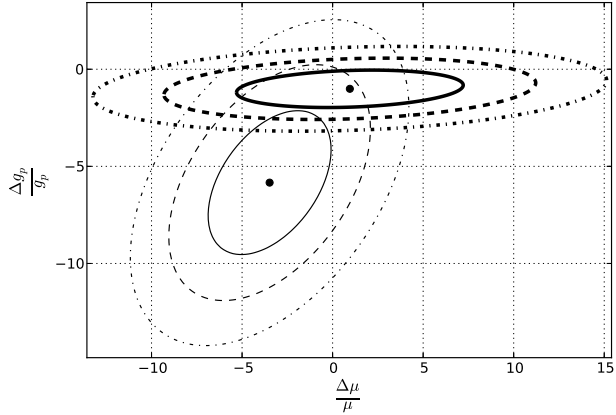}
\end{center}
\caption{Comparing astrophysical constraints with with atomic clock bounds: Horizontal contours correspond to atomic clock measurements from \protect\cite{Clocks} (divided by $H_0$ to make them dimensionless) while tilted contours correspond to the data of Table \protect\ref{table1} plus the $\alpha$ measurement of \protect\cite{LP1}. One, two and three sigma contours are plotted in both cases.}
\label{compclocks}
\end{figure}
%%%%%%%%%%%%%%%%%%%%%%%%%%%%%%%%%%%%%%%%%%%%%%%%%%%%%%%%%
\begin{figure}
\begin{center}
\includegraphics[width=9cm]{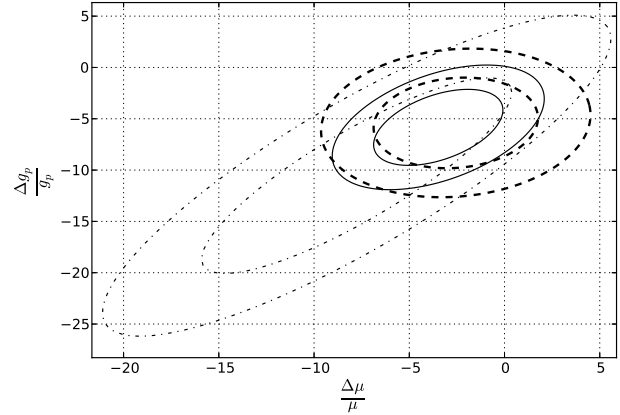}
\end{center}
\caption{One and two sigma constraints on the $\mu$--$g_p$ plane, using the LP $\alpha$ measurement \protect\cite{LP1}. The dashed contours correspond to the low redshift sample (\emph{i.e.}, up to and including the $z=1.336$ measurement); the dash-dotted contours correspond to the high redshift sample (including measurements from $z=1.361$); the solid contours correspond to the full sample. Note the different degeneracy directions at low and high redshift.}
\label{combalpha}
\end{figure}
%%%%%%%%%%%%%%%%%%%%%%%%%%%%%%%%%%%%%%%%%%%%%%%%%%%%%%%%%
\begin{figure}
\begin{center}
\includegraphics[width=9cm]{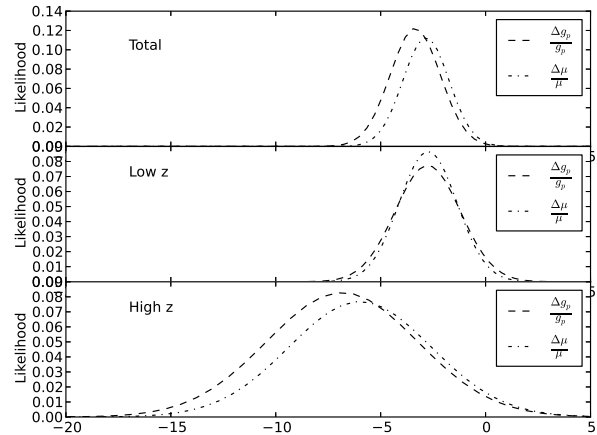}
\end{center}
\caption{One dimensional likelihood for $\mu$ and $g_p$ (marginalizing over the other and using the Large Program $\alpha$ measurement \protect\cite{LP1}), for the full sample as well as for the low-redshift and high-redshift subsamples.}
\label{alphalike1}
\end{figure}
%%%%%%%%%%%%%%%%%%%%%%%%%%%%%%%%%%%%%%%%%%%%%%%%%%%%%%%%%

Figure \ref{combalpha1} shows relevant contours in the $\mu-g_p$, using either the best $\alpha$ measurement (by which we mean the one with the smallest uncertainty, that is \cite{alphaMolaro}) or the LP measurement \cite{LP1}: this is useful to provide intuition on how these combined analyses depend on the choice of (in this case) $\alpha$. As expected, having an $\alpha$ measurement with a smaller uncertainty leads to tighter constraints on the individual parameters. We will now use the LP result as the 'baseline' measurement.

Note also that the degeneracy direction in this plane is different from that coming from current atomic clock measurements; this is illustrated in Fig. \ref{compclocks}. In order to make this comparison we must make the atomic clock results dimensionless, and the simplest way to achieve this is to multiply them by the present Hubble time, $H_0^{-1}\sim1.4\times10^{10}\,yr$. Thus the plot effectively compares astrophysical measurements of $\Delta\alpha/\alpha$ with laboratory measurements of $({\dot\alpha}/H\alpha)_0$. Although the resulting comparison is illuminating, one should bear in mind that no realistic model is expected to have a variation of the couplings that is linear in time from $z=0$ to $z\sim2$, and therefore this rescaling by the Hubble time is somewhat simplistic.

\begin{table}
\begin{tabular}{|c|c|c|}
\hline
 Sample & ${ \Delta\mu}/{\mu}$ (ppm) &  ${ \Delta g_p}/{g_p}$ (ppm)  \\
\hline\hline
$z<1.35$ & $-2.5\pm2.8$ & $-5.4\pm2.9$ \\
\hline
$z>1.35$ & $-7.8\pm5.4$ & $-10.6\pm6.3$ \\
\hline\hline
Full & $-3.5\pm2.2$ & $-5.8\pm2.4$ \\
\hline\hline\hline
\end{tabular}
\caption{\label{marginalpha} 1D confidence intervals listed (at one sigma and in parts per million) for $\mu$ and $g_p$ (marginalizing over the other and assuming the LP $\alpha$ measurement \protect\cite{LP1}), for the full sample as well as for the low-redshift and high-redshift subsamples.}
\end{table}

On the other hand, Fig. \ref{combalpha} displays constraints obtained by dividing the full sample into two equal-sized sub-samples, according to redshift: the transition will thus be at $z=1.35$. As expected, tighter and less degenerate constraints can be obtained at low redshift, but note also that the likelihood contours have different degeneracy directions at high and low redshifts. This stems from the fact that different measurement techniques probe different combinations of couplings, and each technique typically has a limited range of redshifts over which it can be used (due to target availability or other practicalities).

Fig. \ref{alphalike1} shows relevant one-dimensional likelihoods for the full sample and the two subsamples. This leads to the 1D confidence intervals listed (at one sigma) in Table \ref{marginalpha} for the full sample and the subsamples. Interestingly both are non-zero (and negative) at about the one to two sigma level.

These results can be compared with those found in the previous subsection, and the 1D constraints on $\mu$ can also be checked for consistency against the direct $\mu$ measurements in Table \ref{table3}. One finds that in the case of $\mu$ there is no strong disagreement, except in the case of the measurement in \cite{BagdonaiteNew} which is a strong detection with opposite sign. On the other hand, there is some tension between the $g_p$ values in Table \ref{marginalpha} and those inferred in the analysis in section IV.A: the analysis of the present section indicates a negative variation of $g_p$ at more than two sigma.

%%%%%%%%%%%%%%%%%%%%%%%%%%%%%%%%%%%%%%%%%%%%%%%%%%%%%%%%%
\begin{figure}
\begin{center}
\includegraphics[width=9cm]{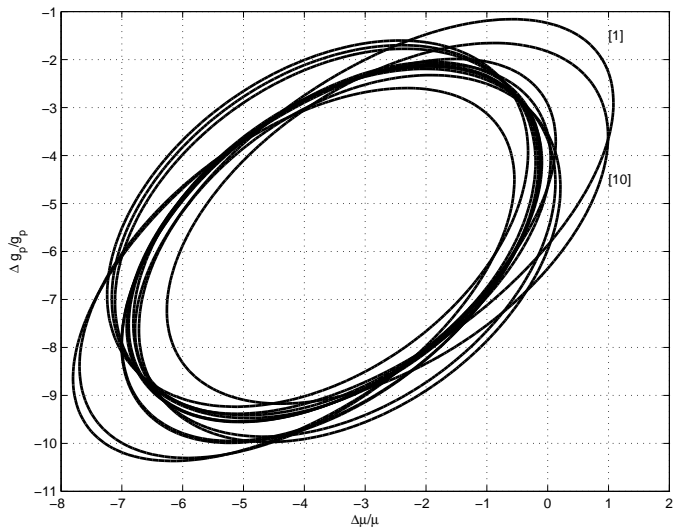}
\end{center}
\caption{Analysis of the data in Table \protect\ref{table1}, assuming the LP $\alpha$ measurement \protect\cite{LP1}. Plotted are the 2D one-sigma likelihood contours, removing one measurement at a time. The number next to each curve identifies which measurement (listed according to the ordering in \protect\ref{table1}) is removed. Clearly, the results are significantly different without \protect\cite{Murphy2} or \protect\cite{Weiss}, and almost unaffected when one of the others is removed.}
\label{jacknife}
\end{figure}
%%%%%%%%%%%%%%%%%%%%%%%%%%%%%%%%%%%%%%%%%%%%%%%%%%%%%%%%%

Naturally, most of the signal for the above detections comes from the two measurements in \cite{Murphy2,Weiss}. This can be easily confirmed, as summarized in Fig. \ref{jacknife}: by redoing the analysis while removing one measurement at a time, we see that the results are significantly different (and much closer to null) when either of these measurements is removed, while they change very little if one of the others is removed.

%%%%%%%%%%%%%%%%%%%%%%%%%%%%%%%%%%%%%%%%%%%%%%%%%%%%%%%%%
\begin{figure}
\begin{center}
\includegraphics[width=9cm]{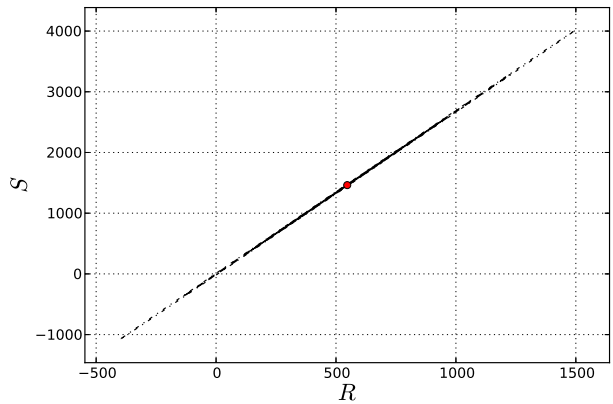}
\includegraphics[width=9cm]{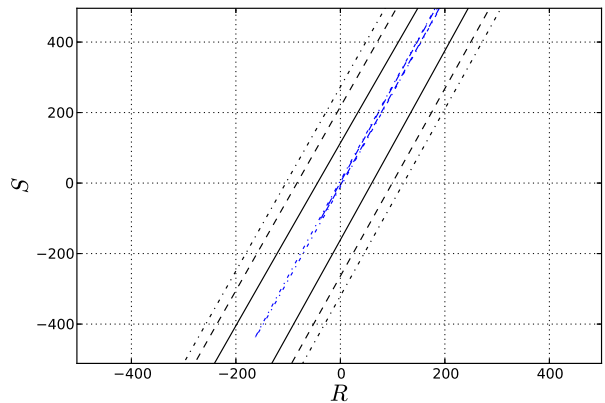}
\end{center}
\caption{{\bf Top panel:} Derived constraints on the $R$--$S$ plane, assuming the $\alpha$ LP measurement in Table \protect\ref{table2}. Solid, dashed and dotted lines correspond to one, two and three sigma contours. {\bf Bottom panel:} Comparing the $R-S$ constraints from astrophysical measurements in the previous panel (narrow central contours) with those obtained from atomic clock data in \protect\cite{Clocks} (wider external contours). Solid, dashed and dotted lines again correspond to one, two and three sigma contours.}
\label{combalpha2}
\end{figure}
%%%%%%%%%%%%%%%%%%%%%%%%%%%%%%%%%%%%%%%%%%%%%%%%%%%%%%%%%

The top panel of fig. \ref{combalpha2} shows the derived constraints on the $R$--$S$ plane, still assuming the $\alpha$ LP measurement in Table \protect\ref{table2}. Comparing this with the corresponding analysis in section IV.A, we notice that although the degeneracy direction is maintained, both the best-fit values and the corresponding uncertainties increase. Specifically, we now find at the one-sigma ($68.3\%$) confidence level
\begin{equation}
R=547\pm411
\end{equation}
\begin{equation}
S=1462\pm1105\,.
\end{equation}
Nevertheless, the two sets of parameters are consistent within their error bars. While the relevance of these numbers is at present unclear given the limitations of the currently available data, these results are encouraging in the sense that they show that this is potentially a very sensitive probe of unification, with may come to fruition with the next generation of facilities. It is also interesting to compare these results with touse obtained in \cite{Clocks} for the atomic clock measurements (at $z=0$); this is done in the bottom panel of fig. \ref{combalpha2}. One notices that the degeneracy direction of both constraints is almost (though not exactly) the same, while the region of the $R-S$ parameter space allowed by atomic clock measurements (which are all null results) is significantly wider.

\subsection{Additional $\mu$ measurements}

A similar analysis can be done with $\mu$: data in Table \ref{table1} can be analyzed assuming one or more of the $\mu$ measurements in Table \ref{table3}. Naturally this will now provide constraints on the $\alpha-g_p$ plane, which can also be compared with the results of the previous subsections, and the $1D$ constraints on $\alpha$ can then be checked for consistency against the direct $\alpha$ measurements in table \ref{table2}, and with the results form the Webb dipole \cite{Dipole}.

%%%%%%%%%%%%%%%%%%%%%%%%%%%%%%%%%%%%%%%%%%%%%%%%%%%%%%%%%
\begin{figure}
\begin{center}
\includegraphics[width=9cm]{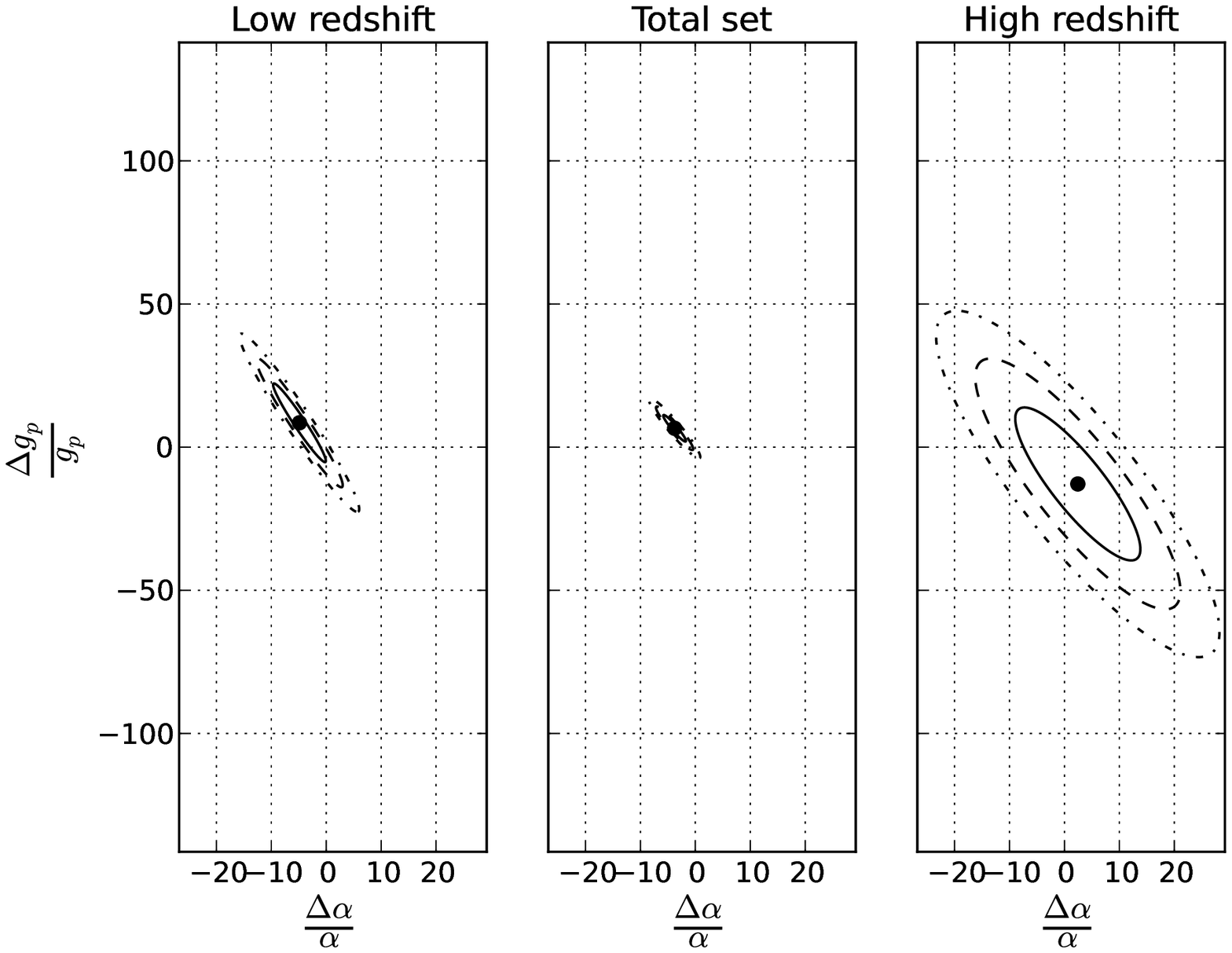}
\includegraphics[width=9cm]{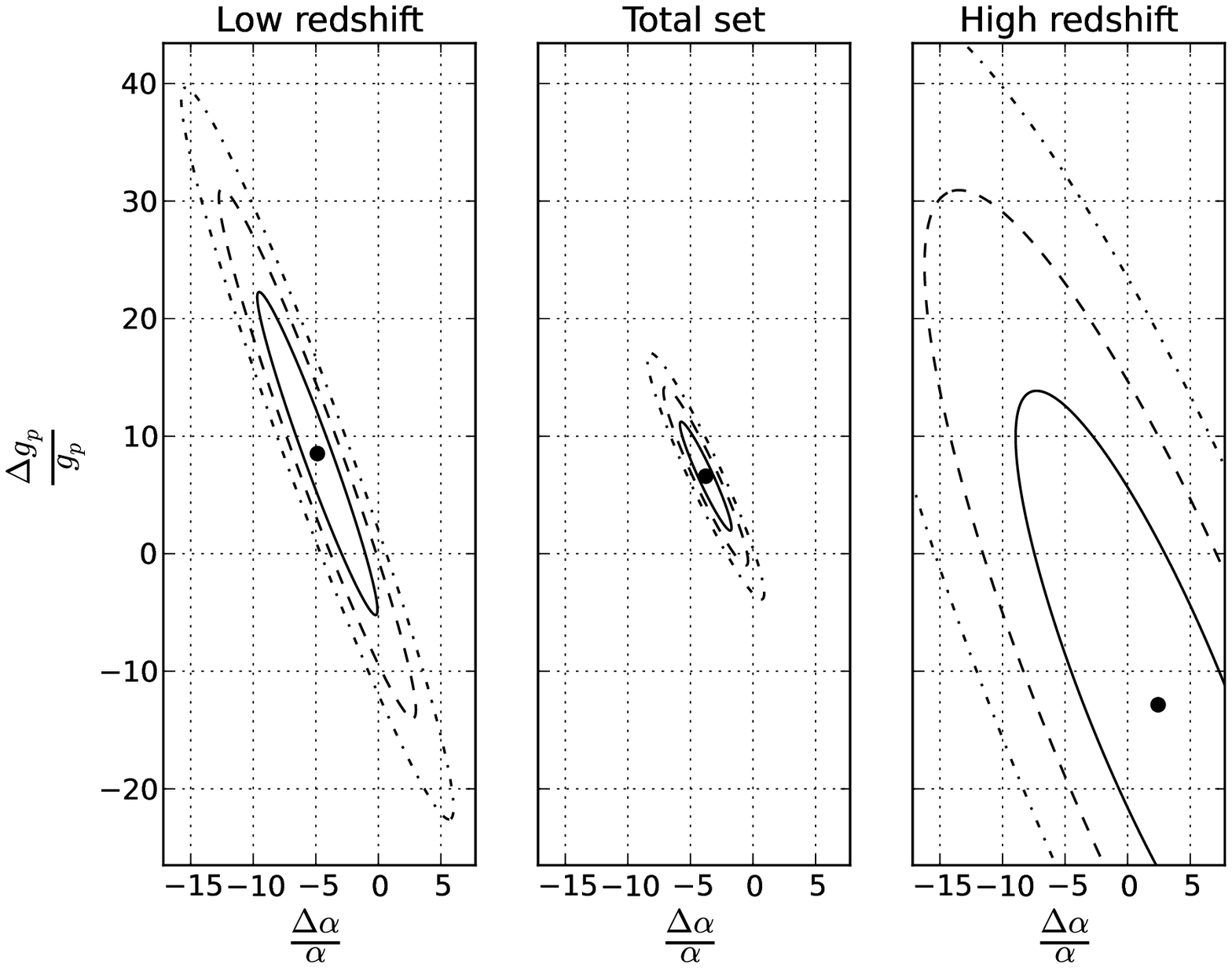}
\end{center}
\caption{Constraints on the $\alpha$--$g_p$ plane, using the \protect\cite{Bagdonaite2} and \protect\cite{LP2} measurements in Table \protect\ref{table3}, respectively for the low and high redshift subsamples, and for the full sample. In all cases, one, two and three sigma contours are shown. The borderline between the two subsamples has been chosen to be at $z=1$. For clarity the bottom panel provides a zoomed version of the top one.}
\label{combmu1}
\end{figure}
%%%%%%%%%%%%%%%%%%%%%%%%%%%%%%%%%%%%%%%%%%%%%%%%%%%%%%%%%
\begin{figure}
\begin{center}
\includegraphics[width=9cm]{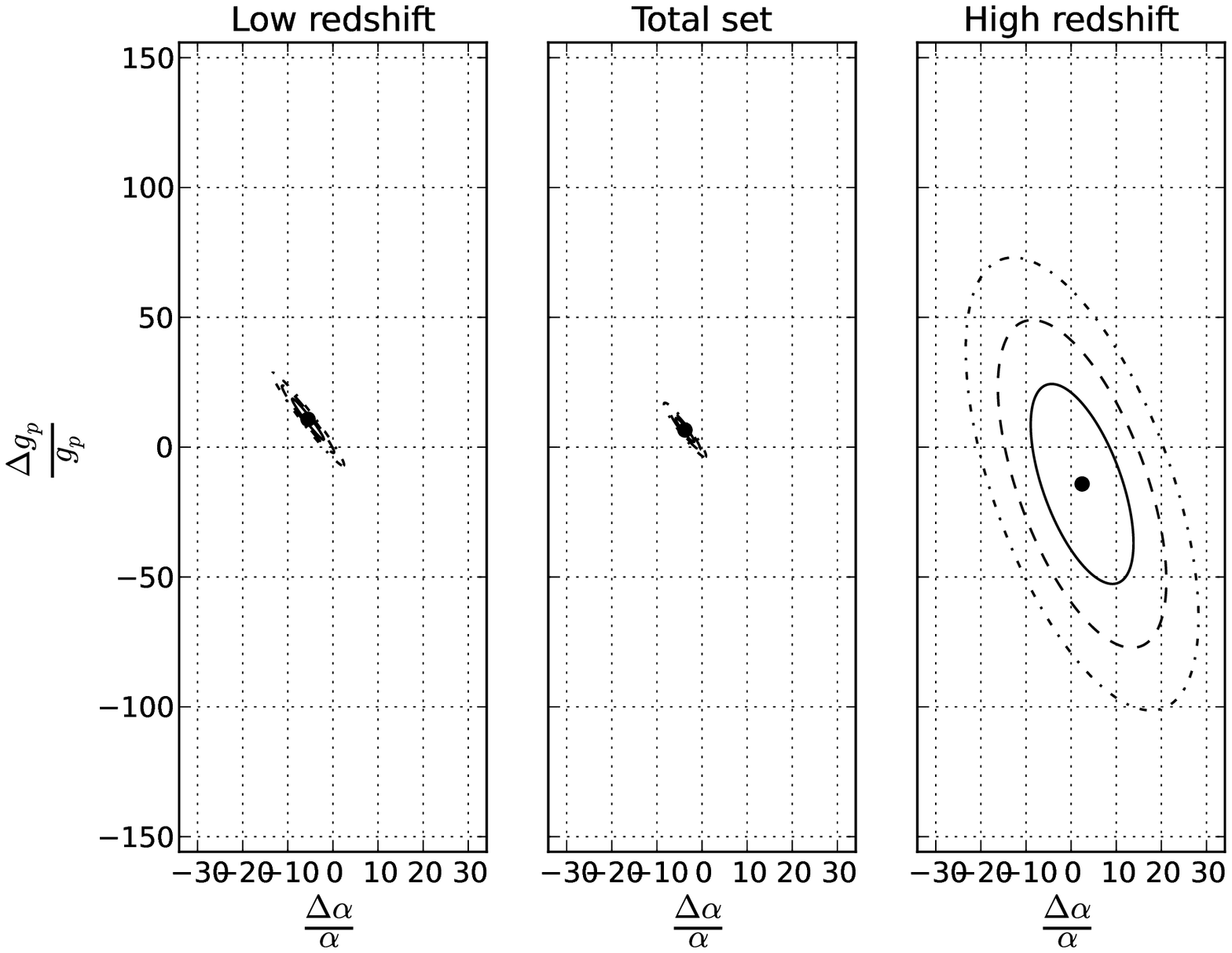}
\includegraphics[width=9cm]{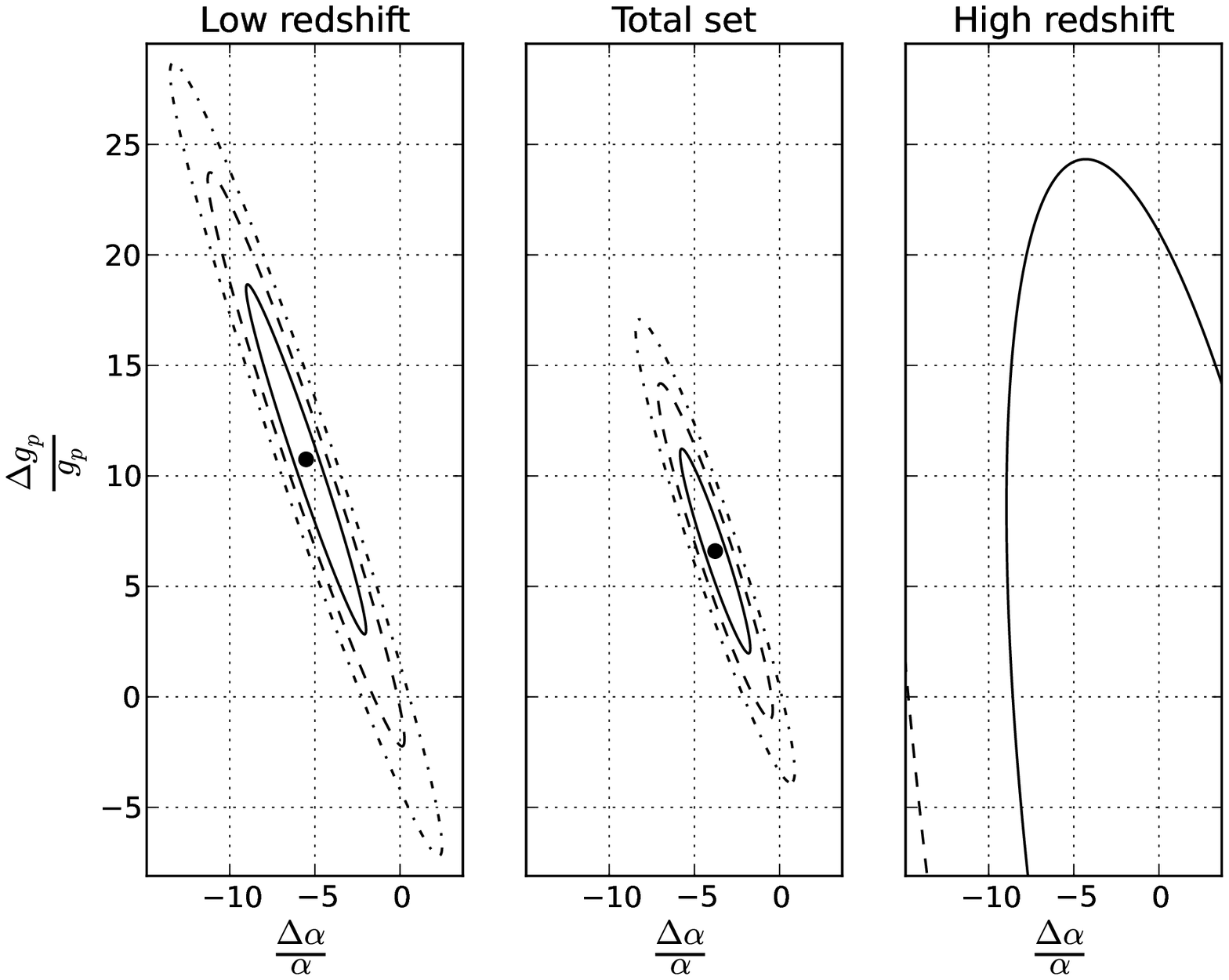}
\end{center}
\caption{As in Fig. \protect\ref{combmu1}, but for a threshold between low and high redshift set at $z=2$ (rather than at $z=1$).}
\label{combmu2}
\end{figure}
%%%%%%%%%%%%%%%%%%%%%%%%%%%%%%%%%%%%%%%%%%%%%%%%%%%%%%%%%

\begin{figure}
\begin{center}
\includegraphics[width=9cm]{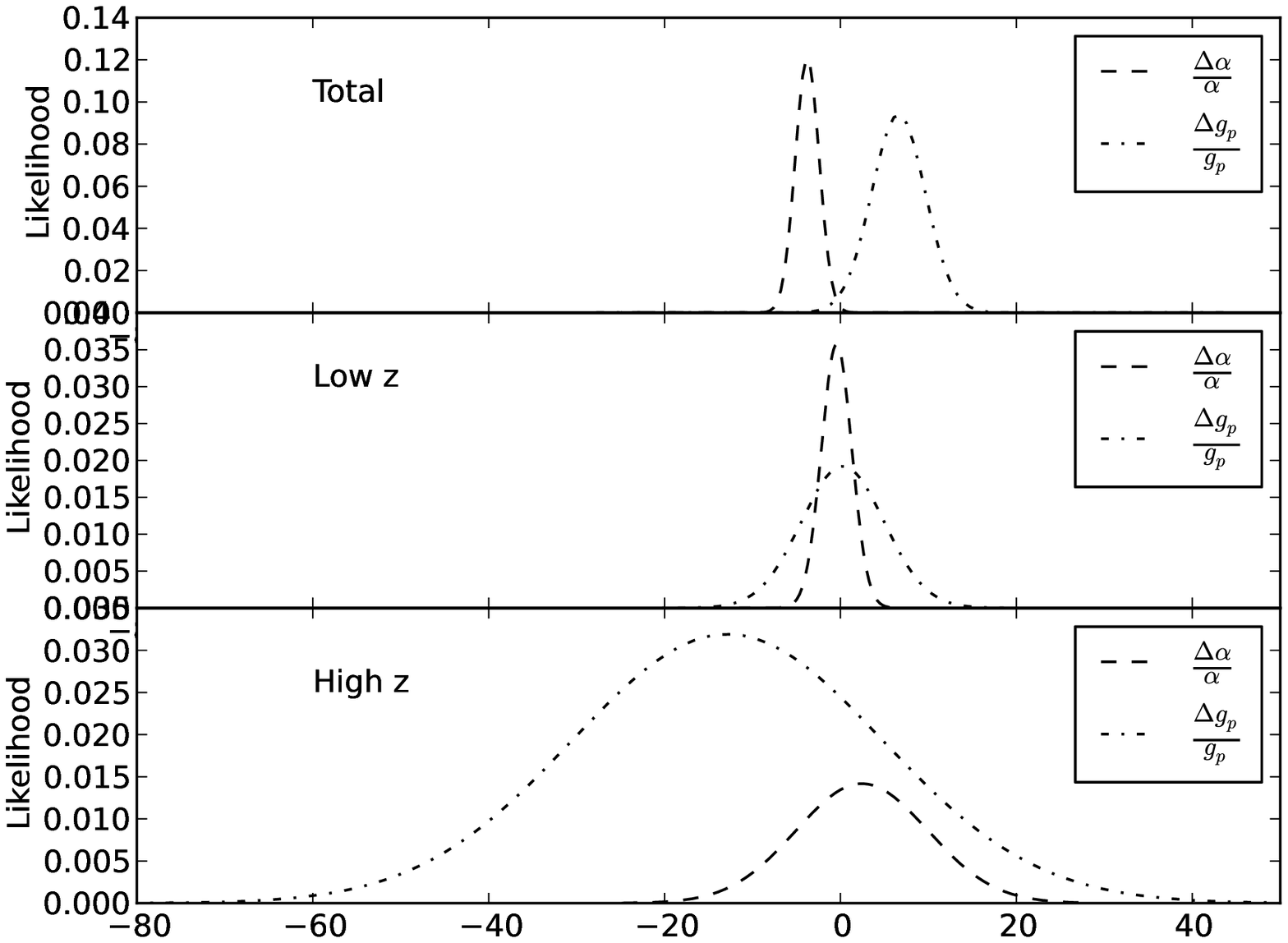}
\includegraphics[width=9cm]{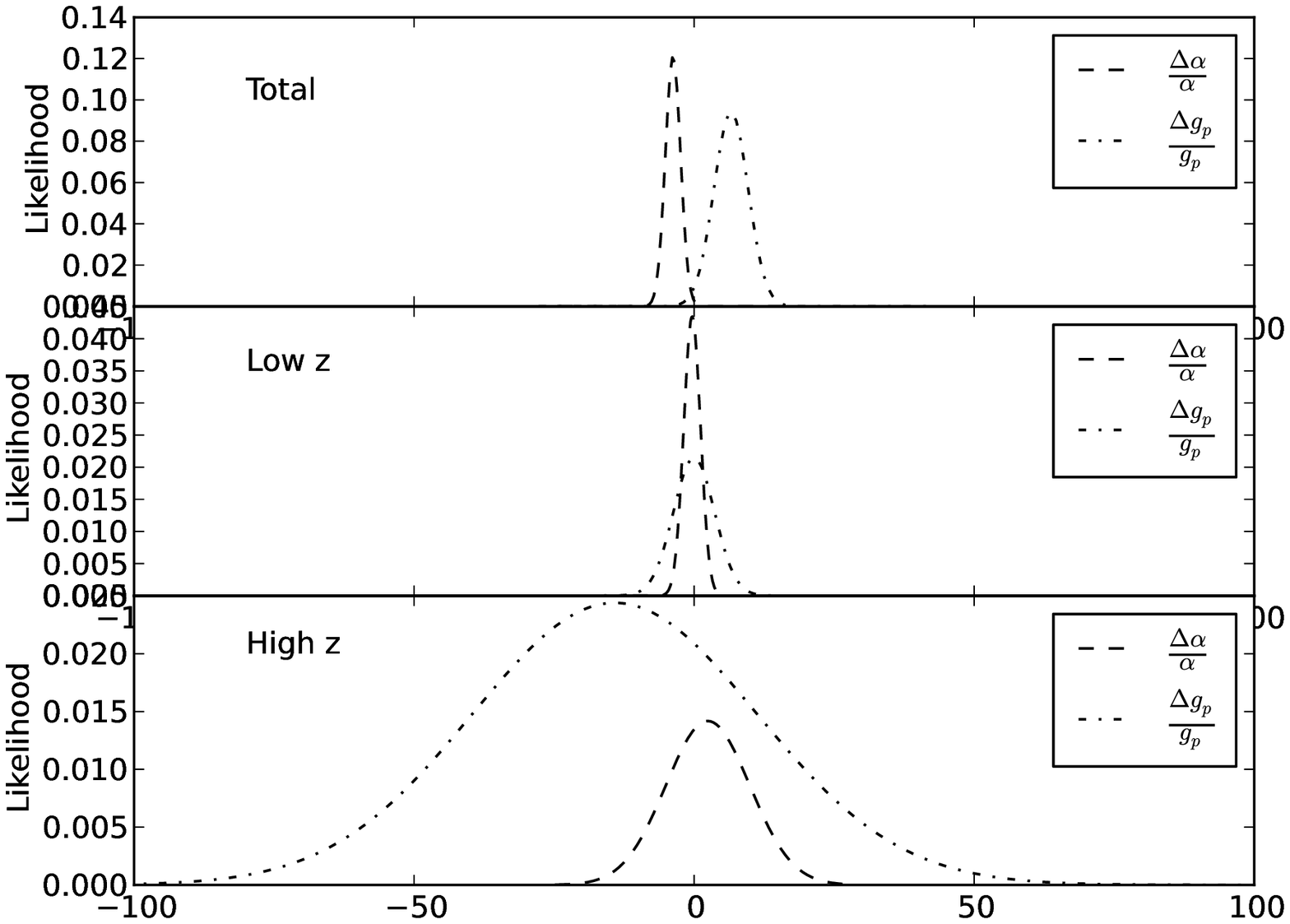}
\end{center}
\caption{One dimensional likelihood for $\alpha$ and $g_p$ (marginalizing over the other and assuming the \protect\cite{Bagdonaite2} and \protect\cite{LP2} measurements in Table \protect\ref{table3} as pivots, respectively for the low and high redshift subsamples), for the full sample as well as for the low-redshift and high-redshift subsamples. In the top panels the division between low and high redshift samples has been as fixed at $z=1$, while in the lower panels $z=2$ has been assumed.}
\label{mulike}
\end{figure}

In this case we will also split the sample into two subsamples according to redshift, using as baseline $\mu$ measurements the Methanol one \cite{Bagdonaite2} for the low redshift subsample and the LP one \cite{LP2} for the high redshift subsample. The results of this analysis are shown in Figs. \ref{combmu1} and \ref{combmu2}. In the former the separation between low and high redshift is set at redshift $z=1$, while in the latter it is set at $z=2$. Here again stronger constraints can be obtained at low redshift. Fig. \ref{mulike} shows relevant 1D likelihoods for the full sample and the two subsamples, while the corresponding 1D confidence intervals are listed (at one sigma) in Table \ref{marginalmu}.

%%%%%%%%%%%%%%%%%%%%%%%%%%%%%%%%%%%%%%%%%%%%%%%%%%%%%%%%%
%%%%%%%%%%%%%%%%%%%%%%%%%%%%%%%%%%%%%%%%%%%%%%%%%%%%%%%%%

\begin{table}
\begin{tabular}{|c|c|c|}
\hline
 Sample & ${ \Delta\alpha}/{\alpha}$ (ppm) &  ${ \Delta g_p}/{g_p}$ (ppm)  \\
\hline\hline
$z<1$ & $-4.9\pm4.8$ & $+8.5\pm13.7$ \\
\hline
$z>1$ & $+2.4\pm11.4$ & $-12.9\pm26.5$ \\
\hline\hline
$z<2$ & $-5.5\pm3.5$ & $+10.7\pm7.9$ \\
\hline
$z>2$ & $+2.4\pm11.4$ & $-14.2\pm38.4$ \\
\hline\hline
Full & $-3.8\pm2.1$ & $+6.6\pm4.6$ \\
\hline\hline\hline
\end{tabular}
\caption{\label{marginalmu} 1D one-sigma confidence intervals listed for $\alpha$ and $g_p$, marginalizing over the other and assuming the \protect\cite{Bagdonaite2} and \protect\cite{LP2} measurements in Table \protect\ref{table3}, respectively for the low and high redshift subsamples.}
\end{table}

Here the results are somewhat interesting: at low redshifts these datasets prefer a slightly negative value of $\alpha$ and a slightly positive value of $g_p$, although in each case this is a less than two sigma effect. The former is consistent with the analysis of Sect. IV.A and (at two sigma) not inconsistent with the direct measurements in Table \protect\ref{table2}, while the latter is inconsistent with what we found in Sect. IV.B (where a negative variation of $g_p$ was preferred).

%%%%%%%%%%%%%%%%%%%%%%%%%%%%%%%%%%%%%%%%%%%%%%%%%%%%%%%%%
\begin{figure*}
\begin{center}
\includegraphics[width=8cm]{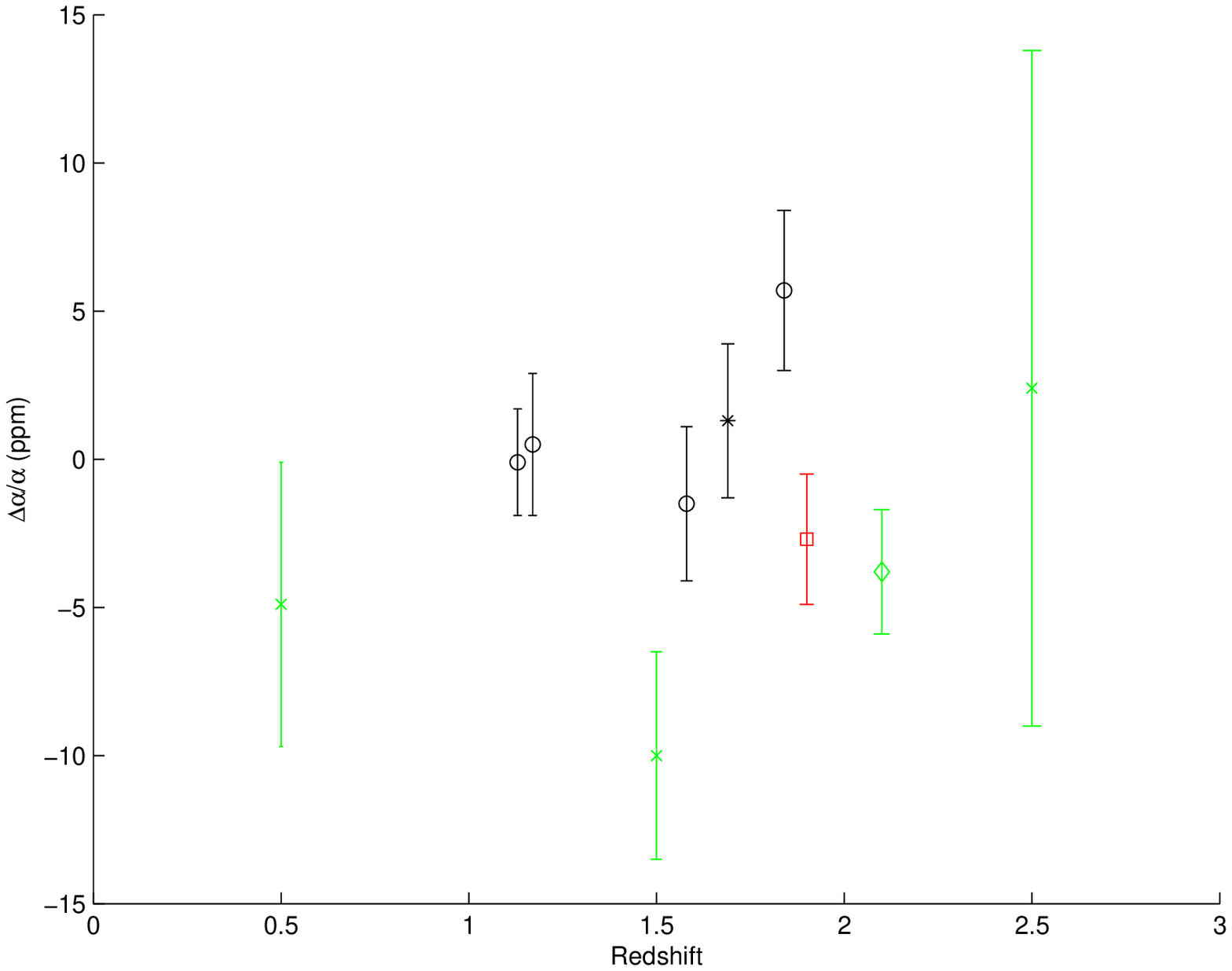}
\includegraphics[width=8cm]{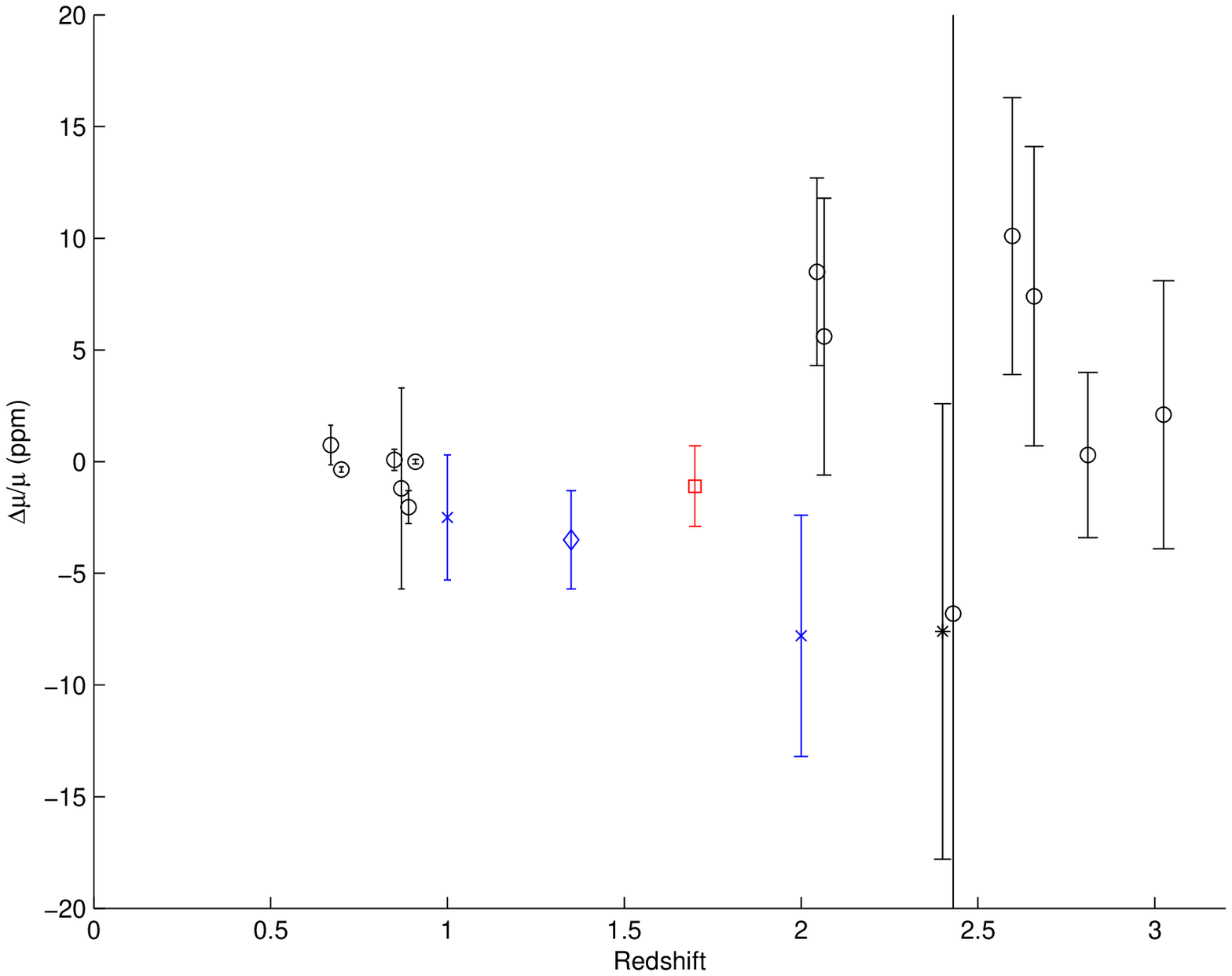}
\includegraphics[width=8cm]{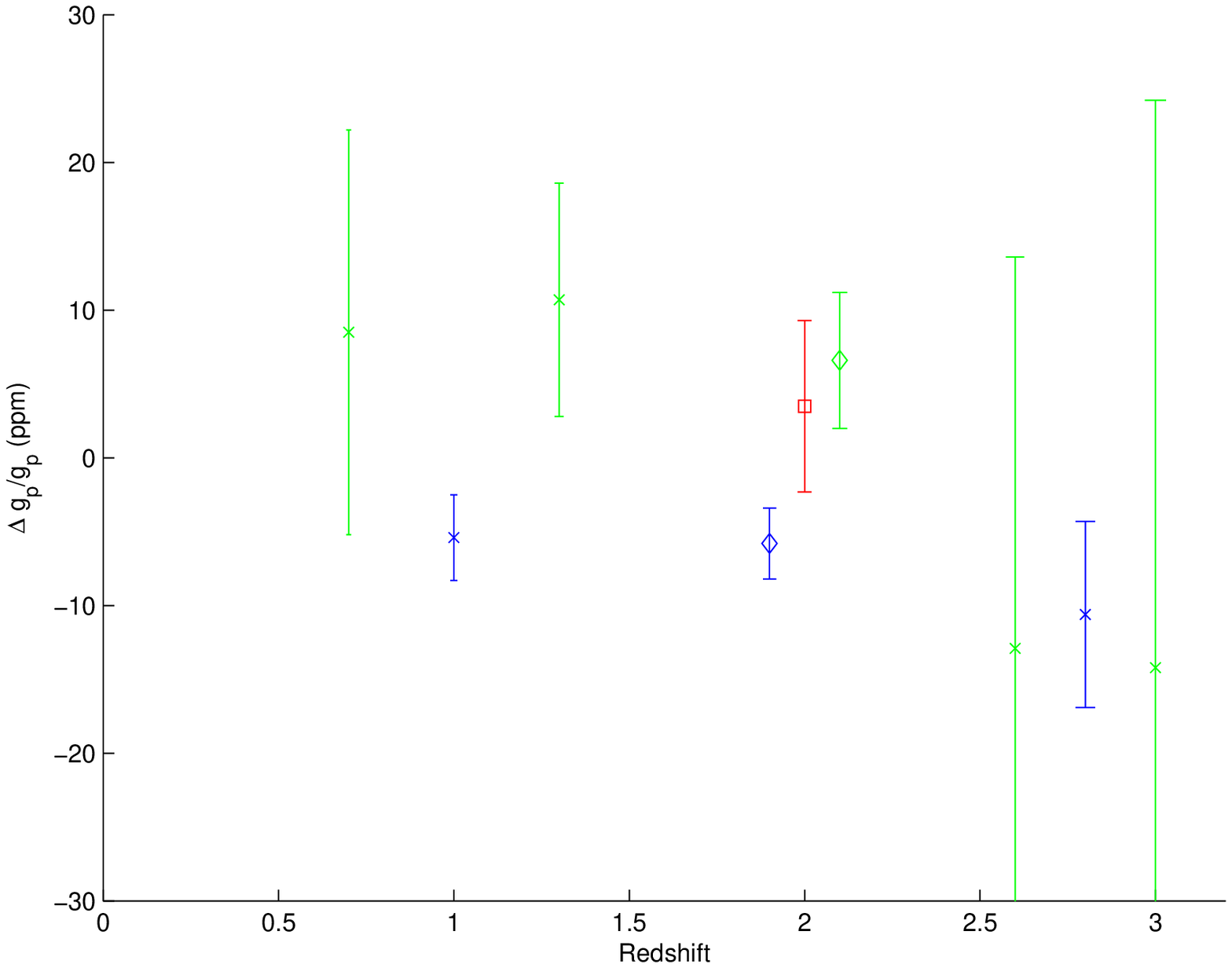}
\end{center}
\caption{Summary comparison of direct and indirect measurements of $\alpha$ (top panel), $\mu$ (middle panel) and $g_p$ (bottom panel). Black circles/asterisks correspond to the {\bf direct} measurements of $\alpha$ and $\mu$ listed in Tables \protect\ref{table2}--\protect\ref{table3}. In both cases the asterisk corresponds to the LP measurement. (Whenever there are several mesurements at the same redshift, these were slightly displaced in redshift to improve readability.) The remaining (colored) points depict the {\bf indirect} measurements obtained in our analysis, as follows: red squares correspond to the results of Sect. IV.A, Eqs. (\ref{fullt1}-\ref{fullt3}); blue points correspond to the results of Sect. IV.B, Table \ref{marginalpha} (the diamond is the full sample result, while the crosses are for the low and high redshift subsamples); the green points correspond to the results of Sect. IV.C, Table \ref{marginalmu} (the diamond is the full sample result, while the crosses are the low and high redshift subsamples). Note that for all indirect results the placement of the points in redshift is purely indicative, as these are the results of combined direct measurements at various redshifts. }
\label{summaryfig}
\end{figure*}
%%%%%%%%%%%%%%%%%%%%%%%%%%%%%%%%%%%%%%%%%%%%%%%%%%%%%%%%%

\subsection{Comparing the various analyses}

Fig. \ref{summaryfig} presents a brief visual summary of our derived results for $\alpha$, $\mu$ and $g_p$ and compares them with the direct measurements of $\alpha$ and $\mu$. We emphasize that the placing of these inferred measurements at particular redshifts is purely indicative, since they are the result of combining direct measurements at various different redshifts. (Still, this is useful for comparison purposes.)

We note that at low redshifts the indirect determinations of $\alpha$ tend to lead to slightly smaller results than the direct ones, while that is not the case for $\mu$. Somewhat more noticeable is the fact that the various methods we used to obtain bounds on $g_p$---either from the data of Table \ref{table1}, or combining these measurements with direct measurements of $\alpha$ or $\mu$---lead to quite different results.

These inconsistencies suggest that the uncertainties in some of the measurements may be underestimated, possibly due to the presence of hidden systematics. While a discussion of the source of these systematics is beyond the scope of the present work, we refer the reader to two of the recent LP publications \cite{LP1,LP2}, where the issue is discussed in considerable detail.

%%%%%%%%%%%%%%%%%%%%%%%%%%%%%%%%%%%%%%%%%%%%%%%%%%%%%%%%%%%%%%%%%%%%%%%%%%%%%%
\section{\label{concl}Conclusions}

We have carried out some simple consistency tests of various recent astrophysical measurements of dimensionless fundamental couplings. Direct measurements of the fine-structure constant $\alpha$ and the proton-to-electron mass ratio $\mu=m_p/m_e$ can be obtained, mostly in the optical/ultraviolet, from a range of absorption systems, typically above redshift $z\sim1$ and up to $z\sim4$. At lower redshifts, on the other hand, various combinations of $\alpha$, $\mu$ and the proton gyromagnetic ratio $g_p$ can be measured, usually in the radio band. The goal of our analysis was to provide a basic comparison between the two types of measurements.

When attempting such comparisons, previous authors often dealt with combined measurements of $\alpha$, $\mu$ and $g_p$ by assuming that only one of these couplings varies (thus turning the combined measurement into one of the coupling in question). This assumption is not justified, in the obvious sense that it does not hold for the vast majority of realistic models. In our analysis the three couplings were allowed to vary simultaneously, and the datasets were used to obtain the most likely values in this parameter space. This somewhat phenomenological approach allows us to compare measurements at various different redshifts, in an approximate but model-independent way.

Naturally a different approach could be taken: one can always choose a specific model (for which the redshift dependence of the couplings will be fully determined), and then compare measurements at different redshifts in the context of this model. This may potentially lead to tighter contraints, though naturally they will be model-dependent. Given the limitations (in quantity and arguably also quality) of the currently available data, we think that a phenomenological approach is more fruitful.

From our analysis it is clear that there are some apparent inconsistencies in the determinations of the various couplings. This is particularly the case in the various indirect determinations of $g_p$. These results support the expectation that hidden systematics may be affecting some of the measurements. Trying to improve this state of affairs is af course the main goal of the ongoing UVES Large Program for Testing Fundamental Physics \cite{LP1,LP2}.

We must also stress that measurements of various combinations $\alpha$, $\mu$ and $g_p$, as well as individual measurements of $\alpha$ and $\mu$ in the same system, are particularly useful. Any Grand-Unified scenario predicts specific (model-dependent) relations between the variations of $\alpha$, $\mu$ and $g_p$. Thus simultaneous measurements of several of these can provide key consistency tests, which will complement (and in some sense are more fundamental than) those that can be done in particle accelerators. Following previous work \cite{Clocks,PKS}, we have briefly illustrated this for the particular class of models previously considered in  \cite{Coc,Luo}, but with suitable generalizations this may be applicable to any unification scenario which includes varying couplings.

Looking further ahead, our results demonstrate the importance of future more precise astrophysical measurements of the stability of these couplings. These future datasets will also make detailed model comparisons possible, including those allowing for spatial or environmental dependencies which may have distinctive signatures \cite{symmetron}. Fortunately, forthcoming facilities such as ALMA and the ESPRESSO and ELT-HIRES spectrographs (at the VLT and E-ELT, respectively), with higher sensitivity and a better control over possible systematics, will provide a detailed and much more accurate mapping of the behaviour of these couplings up to redshift $z\sim4$---and possibly well beyond.

\begin{acknowledgments} 
We are grateful to Mariana Juli\~ao for useful discussions in the early stages of this work. This work was done in the context of the project PTDC/FIS/111725/2009 from FCT (Portugal). O.F. and J.S. acknowledge financial support from \emph{Programa Joves i Ci\`encia}, funded by Fundaci\'o Catalunya--La Pedrera.  C.J.M. is also supported by an FCT Research Professorship, contract reference IF/00064/2012, funded by FCT/MCTES (Portugal) and POPH/FSE (EC).
\end{acknowledgments}

\bibliography{qso}

\end{document}